\documentclass[letterpaper, 10 pt, conference]{ieeeconf}  

\IEEEoverridecommandlockouts                              
\usepackage{graphicx}      
\usepackage{csquotes}

\usepackage[citestyle=numeric-comp,
 style= ieee,
 dashed=false,
 sorting=none, 
 giveninits=true,
 sortcites=true]{biblatex}
\AtEveryBibitem{
    \clearfield{organization}
	\clearfield{issn}
	\clearfield{isbn}
	\clearfield{archivePrefix}
	\clearfield{arxivId}
	\clearfield{pmid}
	\clearfield{pages}
	\clearfield{doi}
	\clearfield{eprint}
	\clearfield{note}
	\clearfield{month}
	\ifentrytype{online}{}{\ifentrytype{misc}{}{\clearfield{url}}}
	\ifentrytype{book}{\clearfield{doi}}{}
}

\usepackage{subcaption}
\usepackage[ruled,vlined]{algorithm2e}
\usepackage{setspace}

\usepackage[usenames,dvipsnames]{xcolor}

\SetCommentSty{mycommfont}

\SetKwInput{Init}{Initialization}
\SetKwInput{InitOpt}{Init. for opt. conv. time}
\SetKwInput{Input}{Input}
\SetKwInput{Output}{Output}
\SetKwBlock{FEND}{for $k=1,2,\ldots$ each node $i$ does}{}

\SetKwBlock{FT}{at each step $t=1,2,3,\ldots $ do}{}
\SetKwBlock{FA}{each node $i\in \V_a$ does}{}

\usepackage{ntheorem,lipsum}
\theorembodyfont{\itshape}

\newcommand{\blue}{\color{blue} }
\usepackage{amsmath,amsfonts,amssymb}

\newcommand{\norm}[1]{{\left\vert\kern-0.3ex\left\vert #1\right\vert\kern-0.3ex\right\vert}}
\newcommand{\mnorm}[1]{{\left\vert\kern-0.3ex\left\vert\kern-0.3ex\left\vert#1\right\vert\kern-0.3ex\right\vert\kern-0.3ex\right\vert}}

\newcommand{\V}{\mathcal{V}}

\title{\LARGE \bf
 Distributed robust secondary frequency control of inverter-based microgrids under time-varying communication delays}

\author{
Milad Gholami, Gianni Bianchini, Antonio Vicino
\thanks{The authors are with Dipartimento di Ingegneria dell’Informazione
e Scienze Matematiche, Università di Siena, Siena 53100, Italy (emails: {\tt\footnotesize \{gholami, giannibi, vicino\}@diism.unisi.it.})}
\thanks{This work was supported in part by the Italian Ministry for Research
in the framework of the 2017 Program for Research Projects of National
Interest (PRIN) under Grant 2017YKXYXJ.
}
}

\begin{document}

\maketitle
\thispagestyle{empty}
\pagestyle{empty}

\begin{abstract}
	This paper presents a robust secondary control strategy for frequency synchronization and active power sharing for inverter-based microgrids. The problem is addressed in a multi-agent fashion where the local controllers of the distributed generators play the role of agents, and communication is affected by time-varying delays. {The approach is fully distributed and based on a synergic combination of linear consensus and integral sliding-mode control. Lyapunov
analysis is presented to assess the stability properties of the closed loop.} Delay-dependent stability conditions are expressed as a set of linear matrix inequalities whose solution yields appropriate control gains such that frequency restoration is achieved despite delays and active power sharing constraints. Simulations confirm the effectiveness of the proposed control strategy.
\end{abstract}



\section{Introduction}
Microgrids (MGs) are small-scale power grids consisting of localized grouping of heterogeneous renewable Distributed Generators (DGs), storage systems, and loads. MGs operate either in islanded, autonomous mode or connected to the main power grid. The control of an AC MG has been recently standardized into a three-layered, nested, control architecture~\cite{m4}. This structure includes a local Primary Control (PC)
level (droop control and primary stabilization), a centralized
or distributed Secondary Control (SC) level (voltage and
frequency restoration), and a tertiary control level (optimal
operation). Among the three, the frequency SC task is of more practical interest to make the integration of renewable generation compliant in the existing fixed-frequency power transmission paradigm. It is also worth mentioning that temporarily modifying the MG frequency while preserving the power sharing among DGs is also useful to perform seamless transition of the MG from islanded to grid-connected mode~\cite{dsilva}.\\ 
Early SCs were centralized \cite{m4}. These solutions are now discouraged in favour of distributed approaches because they scale better with the MG size, are more robust to failures, and dispense from costly central computing and communication units~\cite{dorfler2016breaking}. Pioneering work that solves the frequency SC problem in a distributed fashion is reported in~\cite{simpson2013}. There, a distributed averaging proportional integral scheme is proposed, and power compensation terms are aimed to meet the desired power sharing among DGs. In contrast with~\cite{simpson2013}, where the SC set-points are assumed known to all the DGs, later \emph{leader-follower} strategies have been proposed. These allow for arbitrary adjusting of the MG frequency in a distributed way by acting on the leader DG only. As an example, \cite{guo2015distributed} proposes a leader-based interaction rule and proves its stability features in the local sense.\vspace{0.1cm}\\
Although the mentioned strategies require a communication infrastructure, network-induced communication delays, which
may degrade the MG performance and even destabilize it,
have seldom been considered. The impact on SC of a {\it constant} identical delay for all communication links is analyzed in \cite{liu2015impact} and \cite{droop_based}. An SC that accounts for {\it time-varying} delay is considered in \cite{lai2019stochastic}. A distributed finite-time control protocol for frequency and voltage restoration under a unique {\em constant} delay affecting the communication network is proposed in \cite{Ning2020}, whereas \cite{Andeotti} only addresses the finite-time voltage control problem.
Along this line, recent works \cite{9862562,9902187,9950466} suggest event-triggered SC approaches with {\it time-varying} communication delays. {Although all works mentioned above are promising, they all rely on perfect knowledge
of the DG mathematical models and power measurements}.\\ 
Thus motivated, this paper proposes a leader-oriented frequency SC strategy {capable of restoring DG frequencies to the desired value and establishing active power sharing accuracy in spite of delays in the communication between DGs and unknown load variations. We design a distributed SC consensus protocol consisting of two distinct terms: a  nonlinear sliding-mode-based discontinuous term and a linear time-delay consensus protocol. The nonlinear part of the proposed scheme is an Integral Sliding Mode (ISM)-based
discontinuous term using only local measurements, which
is employed to suppress the effect of disturbances on the MG emerging behavior. The second term aims to compensate for the unavoidable deviations of the DG output frequencies from the set-point value while preserving the power sharing accuracy condition.}  In comparison with \cite{liu2015impact,droop_based,lai2019stochastic,9862562,9902187,9950466,Ning2020}, the proposed strategy requires neither power measurements nor the use of globally known frequency set-points, thus allowing for SC design independently from the PC loop { (i.e., the frequency restoration approach presented in this work is inherently robust against load variations).} As opposed to \cite{Andeotti}, the frequency restoration problem and achieving active power-sharing are the focus of this paper. {The main contributions of the present
work are as follows: (i) We demonstrate that in the absence of disturbance terms affecting the
MG, the linear time-delay consensus protocol forces the output frequency of each
DG to synchronize with the set-point value despite time-varying communication delays. (ii) We prove that considering the presence of the disturbance terms yields after a finite transient a sliding motion along which the MG exhibits the same trajectories of the disturbance-free linear dynamics under the linear time delay consensus protocol, for which the achievement of frequency restoration and active power sharing accuracy was previously demonstrated. (iii) By employing a Lyapunov-Krasovskii function, we derive a delay-dependent control gain tuning procedure, represented as a set of Linear Matrix Inequalities (LMIs) whose solution allows for finding the appropriate control gains. (iv) Lastly, it is worth mentioning that although the proposed SC has
some discontinuous control components since they appear only in their time derivatives, the actual control inputs are smooth. Therefore, they can safely be
used to feed the inner PCs~\cite{m4}. Preliminary results of this research, limited to the voltage SC design, have been presented in~\cite{gholami2018}.}

The article is structured as follows: some preliminary
notions and notations are recalled in Section~\ref{prel}. Section~\ref{sect2} provides the nonlinear inverter-based MG model for SC purposes. The frequency SC problem, along with the proposed strategy and the main results are illustrated in Section~\ref{sect3}. Computer simulations are discussed in Section~\ref{sect4}. Finally, Section~\ref{sect5} provides concluding remarks.
\section{Mathematical Preliminaries and Notations} \label{prel}
\textbf{Notation:}
	The complex, real, and positive numbers sets are denoted by $\mathbb{C}$, $\mathbb{R}$, and $\mathbb{R}^+$, and $\imath=\sqrt{-1}$. Let $A=[A_{ij}]\in\mathbb{R}^{n\times n}$, its transpose is $A^\intercal$. Assume $A$ symmetric, $A\succ0$ ($A\succeq0$) denotes  $A$ positive (semi-) definite. $I_{n}$ is the $n$-dimensional identity matrix and $1_N$ is a vector with all $N$ elements equal to 1. $\mathrm{card}(\mathcal{X})$ is the cardinality of set $\mathcal{X}$. Finally, the $\text{SIGN}(\cdot)$ operator denotes the discontinuous multi-valued function
{\small \begin{equation*}
	\text{SIGN}(\mathcal{S}_i(t)) \in  \begin{cases} 1 & \text{if } \mathcal{S}_i(t)>0 \\
		[1,-1] & \text{if } \mathcal{S}_i(t)=0\\
		-1 & \text{if } \mathcal{S}_i(t)<0
	\end{cases}
\end{equation*}}
 \textbf{Graph Theory:} A graph $\mathcal{G}_N(\mathcal{V},\mathcal{E})$ is a topological tool used to describe pairwise interactions between  \emph{nodes}, or \emph{agents} in the multi-agent field. The \emph{node set} is \mbox{$\mathcal{V}=\{1,2,\ldots,N\}$}, whereas the \emph{edge set} $\mathcal{E}\subseteq \{\mathcal{V} \times \mathcal{V}\}$ describes the pairwise interactions between nodes. The \emph{adjacency matrix} of $\mathcal{G}_N$ is $\mathcal{A}=[\alpha_{ij}]\in\mathbb{C}^{N\times N}$, and $\alpha_{ij} $ denotes the \emph{edge's weight}. If $(i,j)\notin\mathcal{E}$ then $\alpha_{ij}=0$, otherwise $\alpha_{ij}\neq 0$. A \emph{directed path} is an alternating sequence of nodes and edges with both endpoints of an edge appearing adjacent to it in the sequence. Node $i$ is a \emph{root node} if it can be reached from any other vertex by traversing a directed path. 
 \begin{figure}[!h]
	\centering	\includegraphics[width=10pc]{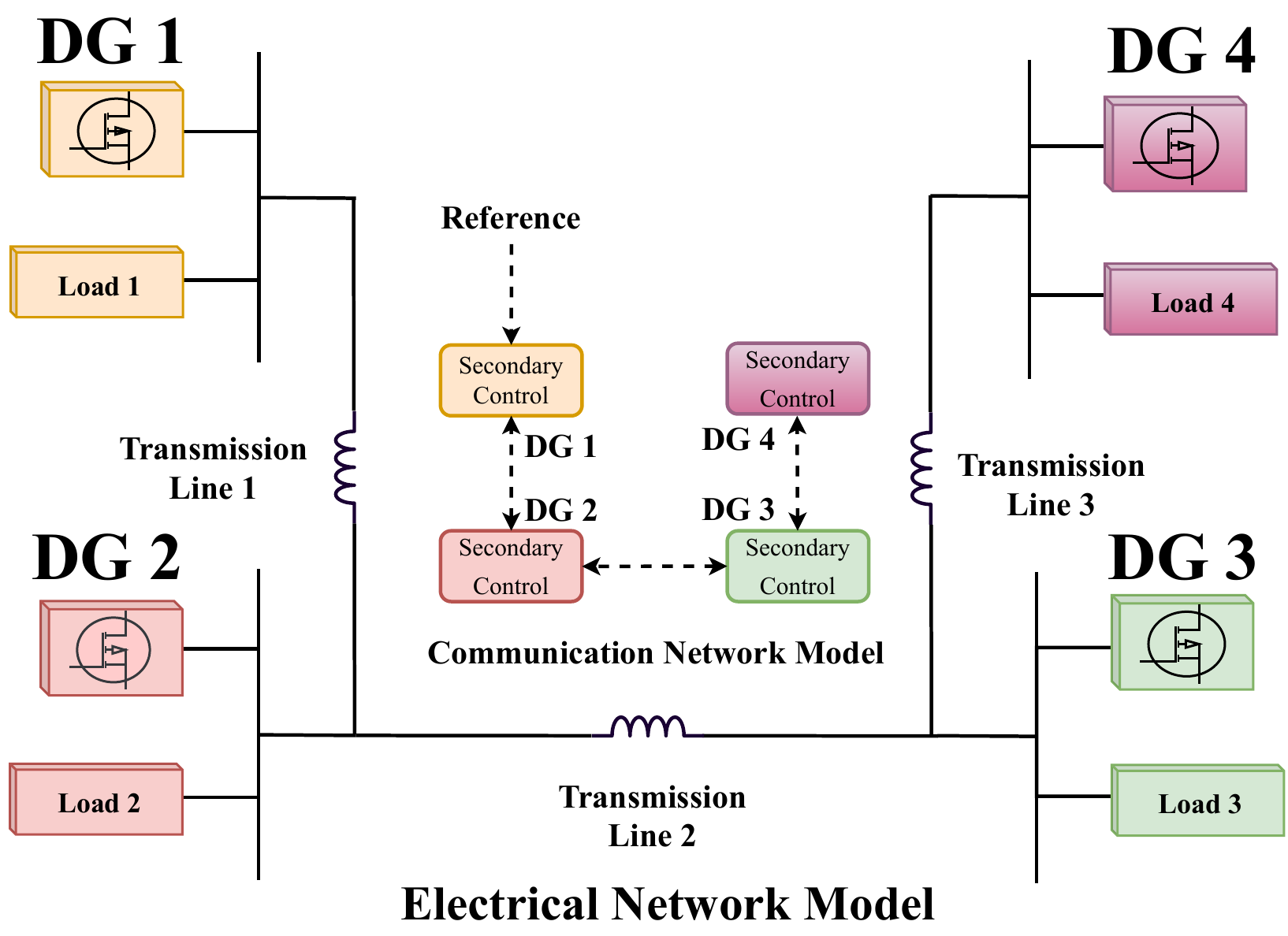}
 \includegraphics[width=10pc]{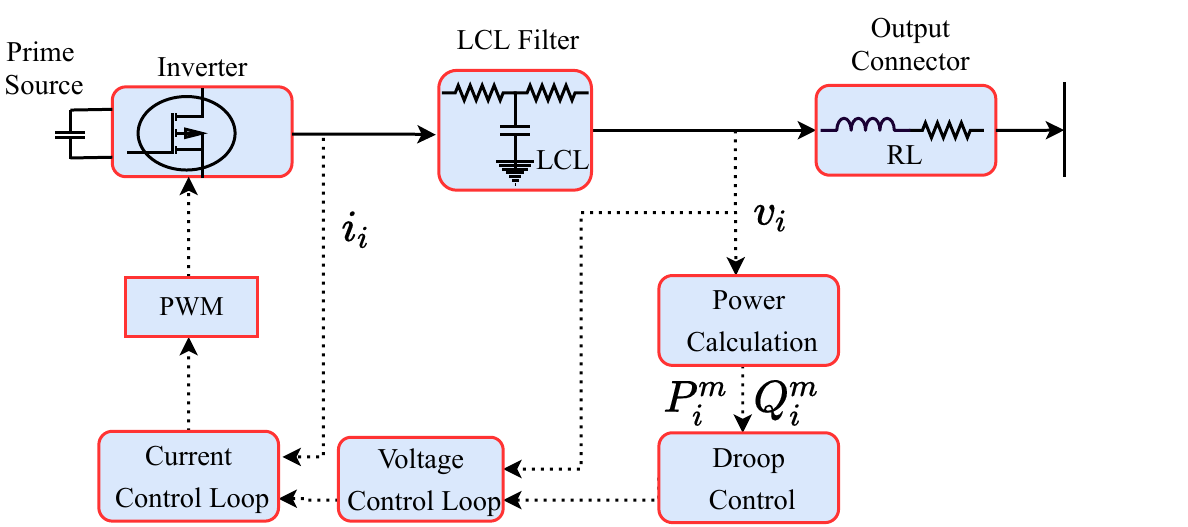}
	\caption{(Left) Cyber and physical layers of a MG composed of four DGs and
		loads; (Right) Block diagram of the PC.\label{fig:DG}}
\end{figure}
\section{Microgrid modeling for SC design}
\label{sect2}
A MG is a cyber-physical system with deeply intertwined components. Interactions among DGs at the cyber and electrical layers can thus be encoded using Graph Theory (see Fig.~\ref{fig:DG}-Left).
A DG includes a DC source, a DC/AC inverter, a smoothing filter, a bus connector, and the PC, which embeds current, voltage, and droop control (see Fig.~\ref{fig:DG}-Right). Its aim is to enforce MG stability under the desired power sharing among DGs. By applying the separation principle, the inner current and voltage PC loops can be neglected \cite{pilloniTIE}, thus yielding the following simplified DG model \cite{guo2015distributed}
\begin{align}
	{  \dot \delta _i(t)=} &{  ~ \omega _i(t)=\bar{\omega}_i(t)-k_{P_i} P_{i}(t) },\quad \forall i\in \mathcal{V}\label{eqn:1}\\
	{   k_{v_i}\dot v_i(t)= }& {  -v_i(t)
		+\bar{v}_{i}(t)-k_{Q_i} Q_{i}(t)},\label{eqn:2}
\end{align}
where $\delta_i$ is the $i$-th DG voltage's phase angle, $\omega_i$ its frequency in rad/sec, and $v_i$ its magnitude in $\mathrm{V_{rms}}$. The frequency and voltage SC inputs are $\bar{\omega}_i$ and $\bar{v}_{i}$, respectively, $k_{v_i}>0$ is the voltage gain, and $k_{P_i}$ and $k_{Q_i}>0$ are the so-called \emph{droop coefficients} selected to meet the given real and reactive power sharing specifications. $P_i(t)$ and $Q_i(t)$ represent active and reactive powers at the DG output ports, dependent on power transmission lines ($P_{ij}$ and $Q_{ij}$) and absorbed powers ( $P_{i}^L$ and $Q_{i}^L$) at load $i$. {\blue 
For more details on the active and reactive power equations, we refer to \cite{gholami2018}}. The electrical interaction among DGs over the MG can be encoded by a complex-weighted connected graph $\mathcal{G}^e(\mathcal{V},\mathcal{E}^e)$ being $\mathcal{V}$ the DG set, $\mathcal{E}^e$ the power lines, and $\mathcal{A}^e=[G_{ij}+\imath B_{ij}]\in\mathbb{C}^{N\times N}$ the complex adjacency matrix, where $G_{ij}$ and $B_{ij}$ denote the conductance and the inductive dominant susceptance between two DGs, respectively. If no connection between the $i$-th DG and the $k$-th DG exists, then $G_{ik}=B_{ik}=0$. Let $\mathcal{N}_{i}^e=\{k:k\in \mathcal{V}, k\neq i, B_{ik}\neq 0\}$. Finally, 
it is worth mentioning that if the SC is inactive, then $\bar{\omega}_i$ and $\bar{v}_{i}$ in \eqref{eqn:1}-\eqref{eqn:2} correspond to the nominal MG utility frequency and rated voltage, e.g., $2\pi50\mathrm{Hz}$ and $220\mathrm{V_{rms}}$.
In the sequel, the following reasonable assumption is made:
\\
\textbf{Assumption 1}\label{ass1}
{\it The actual real and reactive power flows at the DG's output-ports in \eqref{eqn:1}-\eqref{eqn:2} are assumed bounded-in-magnitude by a-priori known constants as
	$ |P_i(t)|\leq \Pi^P_i~, |Q_i(t)|\leq \Pi^Q_i, ~\forall~i\in\mathcal{V}.$}\\
\textbf{Remark 1} {\it Assumption~1 is justified because the power flowing in the lines and/or absorbed by the load is bounded everywhere due to: a) the passive behaviour of loads and lines; b) the bounded operating range of the DC-AC power-converters due to their physical limits; c) the presence of protection apparatus and inner voltage and current PC. This keeps the power flows within pre-specified ranges as discussed, e.g., in \cite{pilloniTIE} and \cite{m3}}.\vspace{0.1cm}\\
Define $d_i(t)$ as the right-hand side of \eqref{eqn:1} without the control term, i.e., $d_i(t)=-k_{P_i}{P}_i(t)$. Given Assumption 1 and the stability ensured by the PC, the following holds
\begin{equation}
	\exists ~\Pi\in \mathbb{R}^+~~:~~ |\dot d_i(t)|\leq\Pi,\quad\forall~t\geq 0.\label{eqn:13_Gamma_Pi}
\end{equation}
Thus Eq. \eqref{eqn:1} can be expressed in the next augmented form
\begin{flalign}
	\dot \omega _i(t)= &\dot{\bar{\omega}}_i(t)+\dot d_i(t)\quad \text{with}\quad \dot d_i(t)=-k_{P_i}\dot{P}_i(t) \label{eqn:dot_omega_i}
\end{flalign}
where $\dot{\bar{\omega}}_{i}$ is the actual control that needs to be designed, and $\dot d_i$ plays the rule of a time-varying disturbance term. \vspace{0.1cm}\\
\textbf{Remark 2}{ \em It is worth remarking that the upper bound \eqref{eqn:13_Gamma_Pi} on signals $\dot d_i(t)$ can easily be estimated
by means of a terminal sliding-mode disturbance observer implemented within each local controller,  see, e.g., \cite{shao2021leakage} and \cite{mobayen2017nonsingular}}.\vspace{0.1cm}\\
\textbf{Cyber MG model:} To achieve the SC objectives in a distributed way, DGs are assumed to be provided with computing and communication facilities embedded within their controllers, which play the role of \emph{agents}. Agents communicate over a network whose topology is encoded by a real-weighted graph $\mathcal{G}^c(\mathcal{V},\mathcal{E}^c)$, where $\mathcal{V}$ is the same node set used to model electrical network $\mathcal{G}^e(\mathcal{V},\mathcal{E}^e)$, whereas $\mathcal{E}^c$ is the set of the available communication links. The set of communication neighbors of the $i$-th DG is $\mathcal{N}_{i}^c=\{j\in \mathcal{V}:~(i,j) \in \mathcal{E}^c\}$. According to the leader-follower paradigm, we further assume the existence of a root node, labeled by $0$ in the augmented graph $\mathbb{G}_{N}^c(\{0\}\cup\mathcal{V},\mathbb{E}^c_{N})$, aimed to provide the SC set-points to a non-empty set of DGs. Clearly, $\mathrm{card}(\mathbb{E}^c_{N})>\mathrm{card}(\mathcal{E}^c)$. Lastly, the set of communication neighbors of the $i$-th DG with respect to graph $\mathbb{G}_{N}^c$ is denoted as $\mathbb{N}_{i}^c=\{j\in \mathcal{V}:~(i,j) \in \mathbb{E}^c_{N}\}$.
\section{Frequency restoration control problem}\label{sect3}
Let the desired frequency set-point from the virtual node $0$ be $\omega_0>0$. In the absence of SCs, with $\bar{\omega}_i=\omega_{0}$ according to \eqref{eqn:1}, the steady-state ($ss$) equilibrium results in $\omega_i(\infty)=\omega_{ss}<\omega_0$, necessitating frequency restoration. As discussed in \cite{simpson2013} and \cite{guo2015distributed} the synchronization condition depends on droop coefficients and the real power balance as 
\begin{equation}
	\omega_{ss} = \omega_{i}(\infty)= \omega_{0}-\frac{\sum_{k=1}^N P_k(\infty)}{\sum_{k=1}^N \frac{1}{k_{P_k}}}\quad\forall~i\in \mathcal{V}
	\label{eqn:omega_ss}
\end{equation}
and preserves the so-called \emph{power-sharing} PC objective
\begin{equation}
	k_{P_i}{P_i(\infty)}=k_{P_k}P_k(\infty) \quad \forall~i,~k\in\mathcal{V}, \label{eqn:sharing}
\end{equation}
whose aim is to enforce a well-defined ratio between the DG injected real power flows. Consider now \eqref{eqn:1}, clearly the achievement of frequency restoration 
\begin{gather} \omega_{ss}=\omega_{i}(\infty)=\omega_{0}\quad \forall~i\in\mathcal{V},
	\label{eqn:freq_sync_control_obj}
\end{gather}
under the constraint~\eqref{eqn:sharing} implies that the frequency SC inputs should be equal to each other at the steady state 
\begin{equation}\label{eqn:consenus_omega_ni}
\bar{\omega}_i(\infty)=\bar{\omega}_{ss}\quad \forall ~i\in\mathcal{V}
\end{equation}
being $\bar{\omega}_{ss}$ the frequency SC input at the steady state.
\section{Frequency secondary control design}\label{sect3}
To solve the SC problem in a distributed setting with network-induced communication delays, we consider a two-component local interaction protocol as follows
\begin{equation}
\dot{\bar{\omega}}_i(t)=\dot u^c_i(t)+\dot u^d_i(t).
\label{udef}
\end{equation}
The first component $\dot u^c_i(t)$ satisfies the linear consensus protocol dynamics
\begin{equation}
	\begin{array}{rl}
		\dot u^c_i(t)=&- \sum_{j\in\mathbb{N}_{i}^c}k_{ij} \left(\omega_{i}(t-\tau(t))-\omega_{j}(t-\tau(t)))\right.\\
		&-\sum_{j\in\mathcal{N}_{i}^c} \overline{k}_{ij} \left(u^c_i(t-\tau(t))-u^c_j(t-\tau(t))\right)
	\end{array}\label{eqn:frequency_SC}
\end{equation}
where ${k_{ij}}$ and $\overline{k}_{ij}$ are scalar control gains to be designed, such that if $(i,j)\in\mathbb{E}^c_{N}$ then ${k_{ij}}>0$ and $\overline{k}_{ij}>0$, otherwise ${k_{ij}}=0$ and $\overline{k}_{ij}=0$. $\tau(t)$ is a uniform time-varying delay associated with communication channels. The common assumption of rate-bounded time-varying delays is made \cite{droop_based}, \cite{Andeotti}, and \cite{fridman2009}.\vspace{0.1cm} \\
\textbf{Assumption 2}{ \em Let known bounds $\tau^{\star}$, $\tau_{g}$, exist, such that}
\begin{align}\label{eqn:ass_delay}
	0\leq\tau(t)\leq \tau^{\star},\quad |\dot{\tau}(t)|\leq \tau_g
	.\end{align}
To correlate the data received from a DG for feedback purposes, we further assume that: \vspace{0.1cm}\\
\textbf{Assumption 3}{ \em $\tau(t)$ is measurable for all $t\geq 0$.}\vspace{0.1cm}\\
\textbf{Remark 3} {\em Communication protocols usually include a data-packet time-stamp. Thus, the requirement for detectable delays is costless. Once the delay $\tau(t)$ is detected, by means of local buffers, each controller is enabled to retrieve its own state at that time $t-\tau(t)$, and performs~\eqref{eqn:frequency_SC}. This assumption is common in networked applications, see, e.g.,~\cite{chen} and \cite{droop_based}.}
\vspace{0.01cm}\\
{The second control component $\dot u_i^{d}(t)$ in \eqref{udef} is the discontinuous nonlinear control
\begin{equation}
\dot u_i^d(t)=-m_i \; \text{SIGN}(\mathcal{S}_i(t)),
\label{usmdef}
\end{equation}
where $m_i$ are scalar control gains and the sliding variable is defined as
$
		\mathcal{S}_i(t)=\omega_i(t)-z_i(t)
$
where $z_i(t)$ satisfies
\begin{equation}
		\dot{z}_i(t)= \dot u_i^c(t),\quad 
		z_i({0})=\omega_i(0).
	\label{sliding3}
\end{equation}
\textbf{Theorem~1}{\label{Thmnew}
Consider the MG dynamics \eqref{eqn:1}-\eqref{eqn:2} under Assumptions 1-3 along with the local interaction rule \eqref{udef}-\eqref{sliding3}. Let the gain parameters $m_i$ be chosen such that
		\begin{equation}\label{tuncond}
		m_i>\Pi.
	\end{equation}
Then, the trajectories of the closed-loop system converge to those of the linear disturbance-free MG dynamics 
\begin{flalign}
	\dot \omega _i(t)= &\dot u^c_i(t) \label{free-dynamic}
\end{flalign}}
\textbf{Proof of Theorem 1.} 
	Consider the Lyapunov function
	\begin{equation}
		\begin{array}{rl}
			V_i(t)=&\frac{1}{2}\mathcal{S}_i(t)^2.
		\end{array}
		\label{slidingly}
	\end{equation}
By straightforward manipulations, we obtain that  along the
	trajectories of \eqref{eqn:dot_omega_i}, \eqref{udef}-\eqref{sliding3} the time derivative of the sliding variable $\mathcal{S}_i(t)$ is given by
 	\begin{equation}\label{Sdot}
  \begin{array}{ll}
       \dot {\mathcal{S}}_i(t)&=\dot \omega_i(t)-\dot z_i(t)=\dot d_i(t)+\dot u_i^d(t) \\
       &=\dot d_{i}(t)-m_i \text{SIGN}(\mathcal{S}_i(t)). 
  \end{array}
\end{equation}
Hence,
	\begin{equation}
		\begin{array}{rl}
			\dot V_i(t)=&\mathcal{S}_i(t)\cdot \dot {\mathcal{S}}_i(t)=\mathcal{S}_i(t)\cdot \dot d_{i}(t)-m_i\cdot|\mathcal{S}_i(t)|.
		\end{array}
		\label{dly}
	\end{equation}
Therefore, by \eqref{eqn:13_Gamma_Pi} and \eqref{tuncond}
	\begin{equation}
		\begin{array}{rl}
			\dot V_i(t)\leq&-|\mathcal{S}_i(t)|\cdot (m_i-\Pi)<0.
		\end{array}
		\label{dly2r}
	\end{equation}
Then by \eqref{sliding3}, $V_i (0) = 0$, which along with \eqref{dly2r} yields that
	\begin{equation}
		V_i(t)=0, \quad \forall t \geq 0 \quad \rightarrow \quad \mathcal S_i(t)=0, \quad \forall t \geq 0,
		\label{dly2}
	\end{equation}
i.e. a sliding motion along the manifold $\mathcal S_i(t)=0$ takes place from the initial time on. According to the equivalent control method for analyzing the sliding-mode dynamics \cite{utkin1996integral}, the trajectories of the discontinuous closed-loop system can be achieved by solving the equivalent closed-loop system where the discontinuous control term $\dot u_i^d(t)$ is substituted by the associated equivalent control $\dot u_{i,eq}^d(t)$. The latter is computed by solving the equation $\dot {\mathcal{S}}_i=0$. It follows from \eqref{Sdot} that
	\begin{equation}
		\dot u_{i,eq}^d(t)=-\dot d_i(t),
		\label{ueq}
	\end{equation}
Replacing \eqref{ueq} for $\dot u_i^d(t)$ in the discontinuous closed-loop dynamics \eqref{eqn:dot_omega_i}, \eqref{udef}-\eqref{sliding3}, we can get the disturbance-free MG dynamics \eqref{free-dynamic}.  $\square$\vspace{0.1cm}\\
The point now is to design $\dot u_i^c(t)$ in \eqref{free-dynamic} to achieve the control objectives \eqref{eqn:freq_sync_control_obj} and \eqref{eqn:consenus_omega_ni} asymptotically. Up to now, we found that \eqref{udef} degenerates into \eqref{free-dynamic} for all $t\geq 0$ according to Theorem 1.} It follows that the achievement \eqref{eqn:freq_sync_control_obj} and \eqref{eqn:consenus_omega_ni} along the trajectories of \eqref{udef}, could
be met if and only if the synchronization error vector
	\begin{equation}
		e(t)=\begin{pmatrix} e_1(t),\dots,e_{N}(t)\end{pmatrix}^\intercal\quad\mathrm{with}\quad e_i(t)=\omega_i(t)-\omega_0
		\label{eqn:e_vec}\end{equation}
	and the disagreement vector associated with $ u^c_i(t)$, $\forall i \in \mathcal{V}$
	\begin{equation}
	\varepsilon(t)=\begin{pmatrix}\varepsilon_1(t),&\dots,\varepsilon_{N}(t)\end{pmatrix}^\intercal=\Omega \begin{pmatrix}  u^c_1(t),\dots,u^c_N(t)\end{pmatrix}^\intercal
		\label{eqn:eps_vec}
	\end{equation}
	go to zero, where $\Omega=I_{N}-{1_N 1_N^\intercal}/{N}$ is the so-called average disagreement matrix, which satisfies
	$$\Omega=\Omega^\intercal\quad,\quad 1_N^\intercal \Omega=0_{N}^\intercal.$$ 
	Indeed, one may observe that $\varepsilon_i=0$ $\forall$ $i$ implies $ u^c_i(t)= u^c_j(t)$ $\forall$ $i,j\in \mathcal{V}$. \vspace{0.1cm}\\
Let us now compute the compact form of the error
dynamics on the sliding manifold by substituting \eqref{eqn:frequency_SC} into \eqref{eqn:dot_omega_i} as
	{\begin{equation}
		\begin{array}{rl}
			\dot \omega_i(t)=&-\sum_{j\in\mathbb{N}_{i}^c}{k}_{ij}\left(\omega_{i}(t-\tau(t))-\omega_{j}(t-\tau(t))\right)\\&-\sum_{j\in\mathcal{N}_{i}^c}\overline{k}_{ij}\left( u^c_i(t-\tau(t))- u^c_j (t-\tau(t)\right)
		\end{array}
		\label{eqn:closed_loop}
	\end{equation}}
Then by differentiating $e_i(t)$ in \eqref{eqn:e_vec} along the trajectories of \eqref{eqn:closed_loop}, one has
\begin{equation}
		\begin{array}{rl}
			\dot{e}_i(t)=&-{k}_{i0} e_{i}(t-\tau(t))\\&-\sum_{j\in\mathcal{N}_{i}^c}{k}_{ij} (e_{i}(t-\tau(t))-e_{j}(t-\tau(t)))\\
			&-\sum_{j\in\mathcal{N}_{i}^c}\overline{k}_{ij} (u^c_{i}(t-\tau(t))-u^c_i(t-\tau(t))).
		\end{array}
	\end{equation}
Moreover, by simple manipulation, the error dynamics associated with the vectors $e(t)$ and $\varepsilon (t)$  can be written as
\begin{equation}\underbrace{\begin{pmatrix} 
\dot{e}(t)\\\dot\varepsilon(t)
\end{pmatrix}}_{\dot\chi(t)}=\underbrace{\begin{pmatrix} 
{K}^c&\overline{K}^c\\ \Omega{K}^c&\Omega\overline{K}^c
\end{pmatrix}}_{A}\underbrace{\begin{pmatrix} 
{e}(t-\tau(t))\\\varepsilon(t-\tau(t))
\end{pmatrix}}_{\chi(t-\tau(t))}\label{eqn:dot_eps1}\end{equation}	
where  
${K}^c=[k_{ij}^c]\in\mathbb{R}^{N\times N}$ and $\overline {K}^c=[\overline{k}_{ij}^c]\in\mathbb{R}^{N\times N}$ have entries as 
		\begin{equation}
				\begin{array}{rl}
					k_{ij}^c=
					\left\lbrace
					\begin{array}{cl}			
						\sum_{j\in\mathbb {N}_{i}^c} k_{ij}  & \mbox{if } ~i= j\\
      	-k_{ij}&   \mbox{if } ~ (i,j)\in\mathcal{E}^c_{N}
     ~   \mbox{and } ~i\neq j\\
     0& \mbox{otherwise}\\
					\end{array}\right.,\\\\
					\overline k_{ij}^c=
					\left\lbrace
					\begin{array}{cl}			
\sum_{j\in\mathcal{N}_{i}^c} \overline k_{ij}  & \mbox{if } ~i= j\\
      	-\overline k_{ij}&   \mbox{if } ~ (i,j)\in\mathcal{E}^c_{N}
     ~   \mbox{and } ~i\neq j\\
     0& \mbox{otherwise}
					\end{array}\right.  , 
				\end{array}\nonumber
			\end{equation}
{ We are now in a position to present sufficient conditions
guaranteeing the global asymptotic stability of the closed-loop
collective error dynamics on the sliding manifold \eqref{eqn:dot_eps1}.\vspace{0.1cm}\\
\textbf{Theorem~2}{  Consider a MG of $\mathcal{V}=\lbrace 1,2,\dots, N\rbrace$ DGs as in \eqref{eqn:1}-\eqref{eqn:2}, under the SC~\eqref{eqn:frequency_SC}. Let the communication topology $\mathcal{G}_{N}^c$ be connected and undirected. Let node $0$ be a root node over $\mathbb{G}_{N}^c(0\cup \mathcal{V}, \mathbb{E}_{N}^c)$. Assume the communications subjected to time-varying delays $\tau$, with given upper-bounds $\tau^\star$ and $\tau_g$ as in~\eqref{eqn:ass_delay}. Let Assumptions 1, 2, and 3 be in force. Let there exist $2N\times2N$ symmetric positive definite matrices $Q$, $R$, $P$, $2N\times2N$ free
matrices $ M$, $T$, $X$, and positive scalars $k_{ij}$ $\forall(i,j)\in\mathbb{E}^c_{N}$ and $\overline{k}_{ij}$ $\forall(i,j)\in\mathcal{E}^c_{N}$, such that the following LMIs hold: }
 \begin{align}
  	\Xi=\begin{pmatrix} 	\Xi_{11}&-{M}^\intercal+{T}&- A-{M}^\intercal+R^\intercal&0_{2N\times2N}\\\star&\Xi_{22}&-{T}^\intercal-R^\intercal&-A^\intercal\\\star&\star&-2{R}&0_{2N\times2N}\\\star&\star&\star&\Xi_{44}\end{pmatrix}
		 \prec 0
			\label{sigmaBar} \end{align}
with 
$
\Xi_{11}=A+A^\intercal+{Q}+{M}^\intercal+{M}$, $
\Xi_{22}=-{Q}(1-\tau_g)-{T}^\intercal-{T}$, $
\Xi_{44}={\tau^{\star}} {P}+X^\intercal+ X
$.
Then the error dynamics on the sliding manifold  \eqref{eqn:dot_eps1} are globally asymptotically stable. \\	
\textbf{Proof of Theorem 2} Pick the  Lyapunov function
\begin{equation}
	\begin{array}{rl}
	{V}(t)&= {V}_1(t)+{V}_2(t)+{V}_3(t)
	\end{array}
	\label{eqn:bar_V10}
\end{equation}
where
\begin{equation}
	\begin{array}{rl}\
		{V}_1(t)&= \chi(t)^\intercal\chi(t)\\
{V}_2(t)&=  \int_{t-\tau(t)}^{t}\chi(s)^\intercal{Q}\chi(t)ds \\
{V}_3(t)&= \int_{-\tau^{\star}}^{0}\int_{t+\theta}^{t}\dot\chi(s)^\intercal  W\dot\chi(s)dsd\theta
	\end{array}
\label{eqn:bar_V10}
\end{equation}
where $W$ is some symmetric positive definite matrix.
Differentiating $ V_1(t)$ along the solution of \eqref{eqn:dot_eps1}, one has
\begin{equation}
	\begin{array}{rl}
		\dot{{V}}_1(t)= &2\chi(t)^\intercal A \chi(t-\tau(t))
	\end{array}
	\label{eqn:bar_V11_new}
\end{equation}
By employing the Newton-Leibnitz formula
$
\chi(t-\tau(t))=\chi(t)-\int_{t-\tau(t)}^{t}\dot \chi(s)ds,
$
Eq. \eqref{eqn:bar_V11_new} can be rewritten as
\begin{equation}
	\begin{array}{rl}
		\dot{{V}}_1(t)= & \chi(t)^\intercal(A+A^\intercal)\chi(t)-2\chi(t)^\intercal A \int_{t-\tau(t)}^{t}\dot \chi(t)ds
	\end{array}	\label{eqn:bar_V11_new_new}
\end{equation}
Taking the time derivative of $V_2(t)$ and $V_3(t)$ along the solution of \eqref{eqn:dot_eps1}, and performing lengthy but straightforward manipulations taking advantage of \eqref{eqn:ass_delay}, it holds
\begin{equation}
		\dot{{V}}_2(t)\leq \chi(t)^\intercal{Q}\chi(t)-\chi(t-\tau(t))^\intercal{Q}(1-\tau_g)\chi(t-\tau(t)),\label{eqn:bar_V13_new}
\end{equation}
\begin{equation}
	\dot{{V}}_3(t)\leq \tau^{\star}\dot \chi(t){W}\dot \chi(t)- \int_{t-\tau(t)}^{t}\dot \chi(s)^\intercal{W}\dot \chi(s)ds.
\label{eqn:bar_V13_new2}
\end{equation}
Now, by applying the Jensen inequality to the integral term in \eqref{eqn:bar_V13_new2}, it yields that
\begin{equation*}
	\begin{array}{rl}
		- \int_{t-\tau(t)}^{t}\dot \chi(s)^\intercal{W}&\dot \chi(s)ds\leq\\&-\big(\int_{t-\tau(t)}^{t}\dot \chi(s)ds\big)^\intercal{W}\big(\int_{t-\tau(t)}^{t}\dot \chi(s)ds\big)
	\end{array}
\end{equation*}
Hence, \eqref{eqn:bar_V13_new2} can be recast as follows:
{\small\begin{equation}
	\begin{array}{rl}
		\dot{{V}}_3(t)\leq \tau^{\star}\dot \chi(t){W}\dot \chi(t)-\big(\int_{t-\tau(t)}^{t}\dot \chi(s)ds\big)^\intercal{W}\big(\int_{t-\tau(t)}^{t}\dot \chi(s)ds\big)	
	\end{array}
	\label{eqn:bar_V13_new_new}
\end{equation}}
and
{\begin{equation}
	\begin{array}{rl}
		\dot{{V}}_3(t)\leq \tau^{\star}\dot \chi(t){W}\dot \chi(t)	
	\end{array}
	\label{eqn:bar_V13_new_new}
\end{equation}}
Finally, summing up \eqref{eqn:bar_V11_new_new}, \eqref{eqn:bar_V13_new} and \eqref{eqn:bar_V13_new_new}, and adding the identically zero terms
{\begin{equation}
	\begin{array}{ll}
		2\big(\chi(t)^\intercal{M}^\intercal+\chi(t-\tau(t))^\intercal{T}^\intercal\big)\times\\\big(\chi(t)-\chi(t-\tau(t))-\int_{t-\tau(t)}^{t}\dot \chi(t)ds\big)
	\end{array}\label{leve2}
\end{equation}
{\begin{equation}
	\begin{array}{ll}
 2\int_{t-\tau(t)}^{t}\dot \chi^\intercal(t)dsR\times\big(\chi(t)-\chi(t-\tau(t))-\int_{t-\tau(t)}^{t}\dot \chi(t)ds\big)
	\end{array}\label{leve2}
\end{equation}    }  
and $2\dot \chi(t)^\intercal F^\intercal\Big(\dot{\chi}(t)- A\chi(t-\tau(t))\Big)$ where $F$ is a nonsingular matrix, we get
\begin{equation}
	\begin{array}{rl}
		\dot{ V}(t)\leq\rho(t)^\intercal\Sigma\rho(t)
	\end{array}
	\label{finalbarv}
\end{equation}
{\small
	\begin{flalign}
		\Sigma=\begin{pmatrix} 	\Xi_{11}&-{M}^\intercal+{T}&- A-{M}^\intercal+R^\intercal&0_{2N\times2N}\\\star&\Xi_{22}&-{T}^\intercal-{R}^\intercal&- A^\intercal F\\\star&\star&-2{R}&0_{2N\times2N}\\\star&\star&\star&{\tau^{\star}} {W}+F^\intercal+F\end{pmatrix}
		\label{sigma_c}\end{flalign}}
  \begin{equation}
	\begin{array}{rl}
		\rho(t)=\begin{pmatrix} \chi(t)^\intercal,\chi(t-\tau(t))^\intercal,\int_{t-\tau(t)}^{t}\dot \chi(s)^\intercal ds,\dot \chi(t)^\intercal\end{pmatrix}^\intercal
	\end{array}\label{rhoc}
\end{equation}
Thus if $\Sigma \prec 0$ is satisfied, then $\dot { V} (t)$ is negative definite. Clearly, $\Sigma\prec 0$ if and only if $\Theta = \Gamma^{-1}\Sigma\Gamma\prec 0$ where $ \Gamma=diag\{I,I,I,F^{-1}\}$. It is easily seen that if \eqref{sigmaBar} holds, then $\Theta\prec 0$ by taking $F=X^{-1}$ and $W=FPF^{-1}$.  Once the LMI problem \eqref{sigmaBar} is solved, the control gains ${k_{ij}}$ and $\overline{k}_{ij}$ for the local interaction rule \eqref{eqn:frequency_SC} are derived.

Once a solution to the LMI problem \eqref{sigmaBar} is found, the control gains ${k_{ij}}$ and $\overline{k}_{ij}$ to be used in the local interaction rule \eqref{eqn:frequency_SC} are derived.  $\square$\vspace{0.1cm}\\
The consequence of Theorems 1 and 2 is that with appropriate $m_i$ gains (according to \eqref{tuncond}) and by solving the LMI problem \eqref{sigmaBar}  for  ${k_{ij}}$ and $\overline{k}_{ij}$, the MG dynamics  \eqref{eqn:1}-\eqref{eqn:2} with Assumptions 1-3 and the local interaction rule \eqref{udef}-\eqref{sliding3} ensure the frequency restoration condition \eqref{eqn:freq_sync_control_obj} and maintain the active power sharing accuracy condition \eqref{eqn:sharing} despite disturbance terms and time-delayed measurements.}
 \section{Numerical simulation}
	\label{sect4}
 The proposed frequency SC is tested 
	on a MG composed of $N=4$ DGs. Its rated voltage of $220\mathrm{V_{rms}}$ per phase at $50\mathrm{Hz}$. The MG is implemented according to \eqref{eqn:1}-\eqref{eqn:2} under the MATLAB/Simulink environment. The \texttt{ode1} solver is used with a sample time of $50\mathrm{\mu s}$. The MG parameters are listed in Table~\ref{table1}. The electrical and communication topologies, $\mathcal{G}^e_N$ and $\mathcal{G}_N^c$, are as in Fig.~\ref{fig:DG}-Left. Only DG 1 can access the SC set-points from virtual node $0$. 
 	\begin{table}[!h]
		\centering
		\caption{Parameters of the MG Test System.\label{table1}}\scalebox{0.87}{
			\begin{tabular}{c|c|c|c|c|c|c|c}
				\hline\hline
				\multicolumn{2}{c|}{DG 1}        & \multicolumn{2}{c|}{DG 2} & \multicolumn{2}{c|}{DG 3} & \multicolumn{2}{c}{DG 4} \\ \hline
				\textbf{$\tau_{P_1}$} & $0.016$ & $\tau_{P_2}$  & $0.016$  & $\tau_{P_3}$  & $0.016$  & $\tau_{P_4}$  & $0.016$  \\ \cline{2-8}
				$\tau_{Q_1}$          & $0.016$ & $\tau_{Q_2}$  & $0.016$  & $\tau_{Q_3}$  & $0.016$  & $\tau_{Q_4}$  & $0.016$  \\ \cline{2-8}
				$k_{P_1}$             &$6e^{-5}$         & $k_{P_2}$     &$3e^{-5}$           & $k_{P_3}$     &$2e^{-5}$           & $k_{P_4}$     &$1.5e^{-5}$           \\ \cline{2-8}
				$k_{Q_1}$             &$4.2e^{-4}$          & $k_{Q_2}$     &$4.2e^{-4}$          & $k_{Q_3}$     &$4.2e^{-4}$          & $k_{Q_4}$     &$4.2e^{-4}$          \\ \cline{2-8}
				$k_{v_1}$             &$1e^{-2}$         & $k_{v_2}$     &$1e^{-2}$           & $k_{v_3}$     &$1e^{-2}$           & $k_{v_4}$     &$1e^{-2}$           \\ \hline\hline
				\multicolumn{8}{l}{Real and reactive local loads - $(\mathrm{W},\mathrm{VAR})$}\\ \hline
				$P_{1}^{\ell_1}$             & $0.01$        & $P_{2}^{\ell_1}$     &$0.01$           & $P_{3}^{\ell_1}$     &$0.01$           & $P_{4}^{\ell_1}$     &$0.01$           \\ \cline{2-8}
				$P_{1}^{\ell_2}$             &$1$         & $P_{2}^{\ell_2}$     &$2$          & $P_{2}^{\ell_3}$     &$3$          & $P_{4}^{\ell_3}$     &$4$          \\ \cline{2-8}
				$P_{1}^{\ell_3}$             &$1e^{4}$         & $P_{2}^{\ell_3}$     &$1e^{4}$          & $P_{3}^{\ell_3}$     &$1e^{4}$          & $P_{4}^{\ell_3}$     &$1e^{4}$          \\ \cline{2-8}
				$Q_{1}^{\ell_1}$             &$0.01$         & $Q_{2}^{\ell_1}$     &$0.01$          & $Q_{3}^{\ell_1}$     &$0.01$          & $Q_{4}^{\ell_1}$     &$0.01$          \\ \cline{2-8}
				$Q_{1}^{\ell_2}$             &$1$         & $Q_{2}^{\ell_2}$     &$2$          & $Q_{3}^{\ell_2}$     &$3$          & $Q_{4}^{\ell_2}$     &$4$          \\ \cline{2-8}
				$Q_{1}^{\ell_3}$             & $1e^4$  & $Q_{2}^{\ell_3}$     & $1e^4$         & $Q_{3}^{\ell_3}$     & $1e^4$         & $Q_{4}^{\ell_3}$     & $1e^4$         \\ \hline\hline
				\multicolumn{2}{c|}{Lines - $[\Omega^{-1}]$} & \multicolumn{2}{c|}{$B_{12}=10$}& \multicolumn{2}{c|}{$B_{23}=10.67$}& \multicolumn{2}{c}{$B_{34}=9.82$}\\ \hline\hline
		\end{tabular}}
	\end{table}
 Communication delays are time-varying and such that, for each oriented link $(i,j)\in\mathbb{E}_{N}^c$, $\dot{\tau}(t)$ is randomly uniformly distributed within $(-1,1)$. The delay bounds in \eqref{eqn:ass_delay} are $\tau=0.5\mathrm{s}$, and $\tau_g=1000$. We selected an upper bound \eqref{eqn:13_Gamma_Pi} on signals $\dot d_i(t)$ as $\Pi=0.05$ by employing the terminal sliding-mode disturbance observer proposed in \cite{shao2021leakage}. Thus the
parameters $m_i = 0.1$ set according to \eqref{tuncond}. To account for a more realistic use case, the presence of the time-delay voltage SC proposed in \cite{gholami2018}, 
is also considered. 
The appropriate control gains in \eqref{eqn:frequency_SC}, achieved by verifying the feasibility problem of LMI
in \eqref{sigmaBar} through Yalmip Toolbox via SeDuMi solver \cite{boyd1994linear}, are
$k_{10}=2.18$, $k_{12}=1.58$, $k_{21}=1.65$, $k_{23}=1.7$, $k_{32}= 1.69$, $k_{34}=1.65$, $k_{43}=1.83$, $\overline{k}_{12}=1.91$, $\overline{k}_{21}=1.65$, $\overline{k}_{23}=1.7$, $\overline{k}_{32}= 1.7$, $\overline{k}_{34}=1.65$, $\overline{k}_{43}=1.83$. To show the robustness features of the proposed SC to different variations on the MG operating working-points, during the simulation, the following events are scheduled:\vspace{0.1cm}\\
	$\bullet$ Event 1 ($t=0\mathrm{s}$): Only the PC is active, with $\bar{\omega}_i(t)=2\pi 50\mathrm{Hz}\approx 314\mathrm{rad/s}$ and $\bar{v}_i(t)=220\mathrm{V_{rms}}$;\\
	$\bullet$ Event 2 ($t=5\mathrm{s}$): The frequency SC loop with $\bar{\omega}_i(t)$ as in \eqref{eqn:frequency_SC} and $\omega_{0}=2\pi50\mathrm{Hz}\approx 314\mathrm{rad/s}$ is closed;\\
	$\bullet$ Event 3 ($t=20\mathrm{s}$): Load 4, i.e. $(P_{4}^{{L}},Q_{4}^{{L}})$ is disconnected;\\
 	$\bullet$ Event 4 ($t=35\mathrm{s}$): Load 4, i.e. $(P_{4}^{{L}},Q_{4}^{{L}})$ is connected;\\
	$\bullet$ Event 5 ($t=40\mathrm{s}$): The voltage SC loop with $\bar{v}_i(t)$ as in \cite{gholami2018} is closed, with $v_{0}=220\mathrm{V_{rms}}$;\\
	$\bullet$ Event 6 ($t=70\mathrm{s}$): The leader reference for the frequency SC is changed from $50\mathrm{Hz}$ to $50.1\rm{Hz}$.\vspace{0.1cm}\\
 \begin{figure}
		\centering
		\includegraphics[width=19pc]{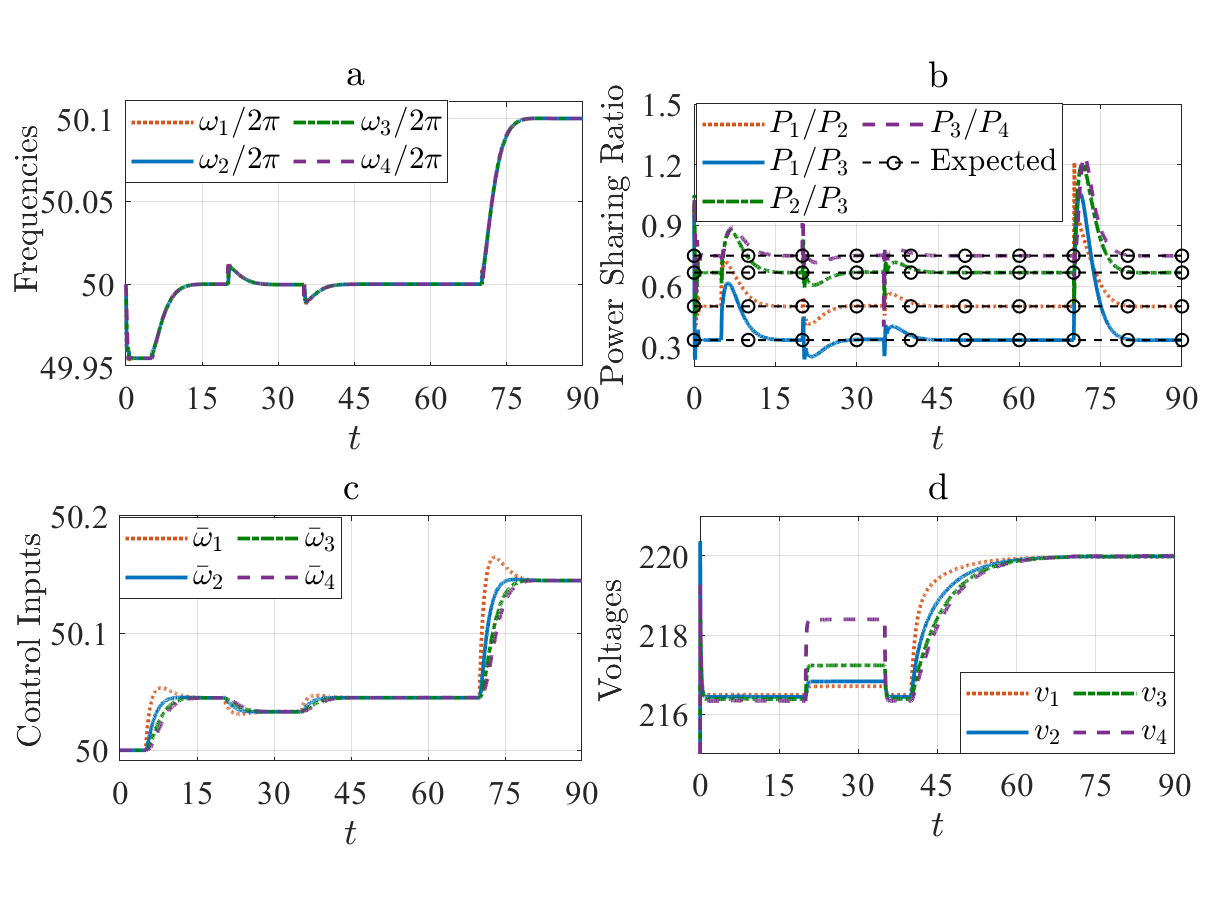}
		\caption{($a$) Output frequencies $\omega_{i}(t)$; ($b$) Comparison between the actual ($P_{i}/P_{j}$) and the expected $k_{P_i}/k_{P_j}$, active power sharing ratio;($c$) Frequency SC control signals $\bar{\omega}_{i}(t)$; ($d$) Output voltages $v_{i}(t)$.		\label{fig_3}}
	\end{figure}
The simulation results in Fig.~\ref {fig_3} confirm that the proposed frequency SC scheme effectively and robustly accomplishes the control objectives \eqref{eqn:sharing} and \eqref{eqn:freq_sync_control_obj} despite the presence of communication time-varying delays, the MG operating working points changes due to either the activation of the voltage SC at $t=40\mathrm{s}$, or the disconnection/connection of the loads, resp., at $t=20\mathrm{s}$, and $t=35\mathrm{s}$. More specifically, when our frequency SC is inactive during Event 1, all the DGs frequencies deviate from their reference values, and the PC cannot succeed in restoring them (see Fig.~\ref{fig_3} (a)). Once activated at $t = 5s$, the proposed distributed SC not only robustly restores the DG frequencies to the expected values (see Fig.~\ref{fig_3} (a)) but also preserves the active power sharing (see Fig.~\ref{fig_3} (b)). Let us further note that, once at $t = 75s$ $\omega_0$ is changed from $50Hz$ to $50.1Hz$, the MG correctly modifies its synchronous frequency while preserving the desired power sharing, cfr. Figs.~\ref{fig_3} (a) and (b). Lastly, it is clear from Fig.~\ref{fig_3} (c) that the frequency SC shows a satisfactory performance and smooth control signals.
	\section{Conclusions}
	\label{sect5}
	This paper has proposed a frequency SC protocol for inverter-based microgrids accounting for disturbance terms and time-varying delays in the communication links. The method improves the current State of the Art because it is fully distributed, model-free, and robust against network-induced time-varying delays. 
	Future research activities could be devoted to considering more general, possibly multi-time varying communication delays and switching topologies.



%

\printbibliography

\end{document}